\documentstyle[aps,prl,floats]{revtex}
\def\note #1]{{\bf #1]}}
\begin{document}
\tighten
\input psfig
\twocolumn[\hsize\textwidth\columnwidth\hsize\csname@twocolumnfalse%
\endcsname
\title{Are Standard Solar Models Reliable?}
\author{John N. Bahcall}
\address{Institute for Advanced Study, Princeton, New Jersey 08540}
\author{M. H. Pinsonneault}
\address{Department of Astronomy, Ohio State University, Columbus,
Ohio 43210}
\author{Sarbani Basu and J. Christensen-Dalsgaard}
\address{Theoretical Astrophysics Center,
Danish National Research Foundation, and \\
Institute for Physics and Astronomy, Aarhus University, DK 8000 Aarhus C,
Denmark}
\date{\today}
\maketitle
\begin{abstract}
The sound speeds of 
solar models that include element diffusion agree with
helioseismological measurements 
to a rms discrepancy of better than
$0.2\%$ throughout almost the entire sun.  Models that do not
include diffusion, or in which the interior of the sun is assumed to be
significantly mixed, are effectively
 ruled out by helioseismology. Standard
solar models predict the measured properties of the sun more accurately
than is required for applications involving solar neutrinos.
\vspace*{.1in}
\end{abstract}

\pacs{PACS numbers:}

]


For almost three decades, a discrepancy has existed between
solar model predictions of neutrino fluxes and the rates observed in
terrestrial experiments.  In recent years,  
the combined results from  four solar
neutrino experiments 
have sharpened the discrepancy in ways that are  independent
of details of the solar models\cite{newphysics}.  
This development is of  broad 
interest since a modest extension of  standard electroweak 
theory, in which neutrinos have small
masses and lepton flavor is not conserved, leads to results in excellent 
agreement with
experiments\cite{MSWetc}. 

Since the implications of a discrepancy with the standard electroweak
model are of great importance, 
 the question persists:  Can the solar neutrino problems
be ``solved'' (or at least  alleviated) by changing the solar
model?  
This question has led to a series of generally unsuccessful
 {\it ad hoc} ``Non-Standard''
solar models\cite{Bahcall89} in which large changes in the physics
 of the
sun are hypothesized in order to lower the calculated rate of the $^8$B 
neutrino
flux. 
Over the past two decades, 
the most often hypothesized change is some form of mixing of the solar
material that reduces the central temperature and therefore the
important $^8$B neutrino flux\cite{EC68,BBU68,SS68,Schatzman69,Press81,R85}.
Previous 
arguments that  extensive mixing does not occur 
are theoretical, including the fact that  
the required energy is five orders of magnitude
larger than the total present rotational 
energy\cite{Bahcall89,R85,Spruit87}.
Most recently, Cumming and Haxton\cite{Haxton96} 
proposed a flow of $^3$He,
characterized by three free parameters,
designed to mix the sun in such a
way as to minimize the discrepancy between
solar neutrino observations and predictions.   
By adjusting the parameters, these authors are able to
reduce the calculated
$^7$Be flux more than the $^8$B flux, a result not achieved in
previous Non-Standard solar models.

The diagnostic power of helioseismology\cite{Basu96a} has been
improved recently  through the 
development by Tomczyk {\it et al.}\cite{Tom95}
of an instrument that  measures 
with the same equipment the 
low- and intermediate-degree mode frequencies.
By providing a consistent set of frequencies for the lowest-degree modes,
which penetrate to the greatest depth in the sun,
these data constrain the properties of the solar core
more tightly than earlier measurements.

In this letter, we compare the solar sound speed $c$ inferred 
from the first year of data\cite{Basu96b}
with sound speeds computed from 
standard solar models   used to predict solar
neutrino fluxes and find a rms   agreement better than  $0.2$\%
over essentially the entire  sun, with {\it no} adjustment of parameters.
Since the deep solar interior behaves essentially
as a fully ionized perfect gas, $c^2 \propto T/\mu$
where $T$ is temperature and $\mu$ is mean molecular weight;
thus even tiny fractional  errors in  the model  
values of $T$ or $\mu$
would produce measurable discrepancies  in the precisely determined
helioseismological sound speed
\begin{equation}
{\delta c \over c}  \simeq 
{1 \over 2} \left({\delta T \over T}  - {\delta \mu \over \mu }\right)
\; .
\label{deltac}
\end{equation}
This remarkable
agreement between standard predictions and helioseismological 
observations rules out   solar models with 
temperature or mean molecular weight 
profiles that differ significantly from standard profiles.
The helioseismological data essentially rule out solar models in which
deep mixing has occured (cf. \cite{els90}) 
and argue against unmixed models in which the
subtle effect of particle diffusion--selective sinking
of heavier species in the sun's gravitational field--is not included.

\begin{figure}[t]
\psfig{figure=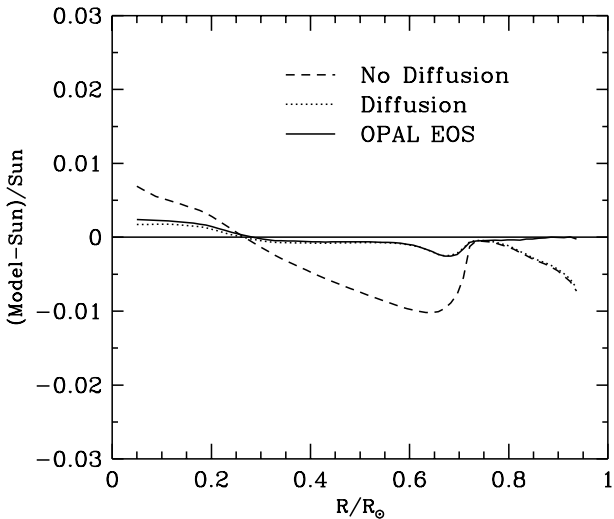,width=3.2truein}
\caption[]{Comparison of sound speeds predicted by different
standard solar models with the
sound speeds measured by  helioseismology.  
There are no free parameters in the models; the microphysics is
successively improved by first including diffusion and then by using
a more comprehensive equation of state.
The figure shows the
fractional difference, $\delta c/c$, 
between the predicted model sound speed and the 
measured\cite{Basu96a,Basu96b}
solar values as a function of radial position in the sun
($R_\odot$ is the solar radius).
The dashed line refers to a model\cite{BP95} in which
diffusion is neglected  and the dotted line was computed
from a model\cite{BP95} in which helium and heavy element diffusion are 
included.
The dark line represents a model which includes 
recent improvements in the 
OPAL equation of state and opacities\cite{Opacity,eos}.\label{fig:one}}
\end{figure}

Figure~\ref{fig:one} compares the  sound speeds computed
from three different solar models with the values
inferred\cite{Basu96a,Basu96b} from the helioseismological measurements.
The 1995 standard model of Bahcall and Pinsonneault (BP)\cite{BP95},
which includes  helium and heavy element diffusion, is represented by
the dotted line; the corresponding BP model 
without diffusion is represented by the dashed
line. The dark line represents the  best solar model which includes
recent improvements\cite{Opacity,eos} in the OPAL equation of state and
opacities, as well as helium and heavy element diffusion.
For the OPAL EOS model,
the rms discrepancy  
between  predicted and measured sound speeds
is  $0.1$\% (which may be due partly to systematic uncertainies in the
data analysis).

In the outer parts of the sun, in the convective region between 
$0.7 R_\odot$ to $0.95 R_\odot$ (where the measurements end), the
No Diffusion and the 1995 Diffusion model have 
discrepancies as large as $0.5$\% (see Figure~\ref{fig:one}).  
The model with the Livermore
equation of state\cite{eos}, OPAL EOS, fits the observations remarkably
well in this region.  We conclude, in agreement with the work of other
authors\cite{Guenther96}, that the OPAL (Livermore
National Laboratory) equation of state provides 
a significant improvement in the description of
the outer regions of the sun.

The agreement between standard models and solar observations
is independent of the finer details of the solar model.  
The standard model of Christensen-Dalsgaard {\it et al.}\cite{science96},
which is derived from an independent computer code with  different
descriptions of the microphysics, 
predicts solar sound speeds that agree everywhere with the measured
speeds to better than $0.2$\%.

Figure~\ref{fig:one} shows that the discrepancies with the No
Diffusion model are as large as 
$1$\%.  
The mean squared discrepancy for the No Diffusion model is 22 times
larger than for the best model with diffusion, OPAL EOS.  If one
supposed optimistically that the No Diffusion model were correct, one
would have to explain why the diffusion model fits the data so much
better. 
On the basis of Figure~\ref{fig:one}, we conclude that otherwise
standard solar
models that do not include diffusion, such  as the model  of 
Turck-Chi\`eze 
and Lopez\cite{TC93}, 
are inconsistent with helioseismological observations.
This conclusion is consistent with earlier inferences based upon 
comparisons with  less complete
helioseismological data\cite{Basu96a,jcd93,els90}, including the fact that
the present-day surface helium abundance in a standard solar 
 model agrees with
observations only if diffusion is included\cite{BP95}.

Equation~\ref{deltac} and Figure~\ref{fig:one} imply that any changes
$\delta T/T$ from the standard model values of  temperature 
must be almost exactly canceled  by 
changes $\delta \mu/\mu$ in mean molecular weight.
In the standard model, $T$ and $\mu$ vary, respectively, by a factor
of $53$ and $43$\% over the entire range for which $c$ has been
measured and by $1.9$ and $39$\% over the energy producing region.
It would be a remarkable coincidence if nature chose $T$ and $\mu$
profiles that individually differ markedly from the standard model but
have the same ratio $T/\mu$.
Thus we expect 
that the fractional differences between the solar and the model
temperature, $\delta T/T$, or mean molecular weights, $\delta \mu/\mu$,
are of similar magnitude to $\delta c^2/c^2$, i.e. (using the larger 
rms error, $0.002$, for the solar interior),

\begin{equation}
\vert \delta T/T \vert, ~\vert \delta \mu/\mu \vert ~ \lesssim ~ 0.004 .
\label{inequality}
\end{equation}

How significant for solar neutrino studies 
is the agreement between observation and prediction
that is shown in Figure~\ref{fig:one}? 
The calculated neutrino fluxes 
depend upon the central
temperature of the solar model 
approximately as a power of the temperature, 
${\rm Flux} \propto T^n$, where  for standard models 
the exponent $n$ varies 
from $n \sim -1.1$ for
the  $p-p$ neutrinos to  $n \sim +24$ for the $^8$B 
neutrinos\cite{BU96}.  
Similar temperature scalings are found for non-standard solar models
\cite{Castellani94}. Thus,
 maximum temperature differences of
$\sim 0.2\%$ would produce changes in the different neutrino
fluxes of several percent or less, much less than 
required\cite{newphysics} to
ameliorate the solar neutrino problems.

\begin{figure}[thb]
\psfig{figure=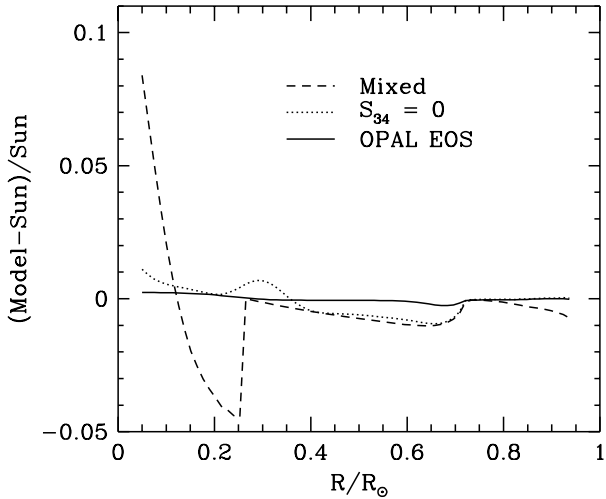,width=3.2truein}
\vskip -0.2in
\hglue-.2in\caption[]{Non-standard solar models compared with 
helioseismology.
This figure is similar to Figure \ref{fig:one} except that the
vertical scale is expanded. The dashed curve represents the sound
speeds computed for the mixed solar model of Cumming and
Haxton\cite{Haxton96}  with
$^3$He mixing.  The dotted line represents the sound speed for a
solar
model computed with the rate of the 
$^3{\rm He}(\alpha,\gamma)^7$Be reaction 
set equal to zero.  
For comparison, we also include the results for the
new standard model labeled OPAL EOS in Figure~\ref{fig:one}.
\label{fig:two}}
\end{figure}

Figure~\ref{fig:two} shows that 
the ``mixed'' model of Cummings and Haxton (CH)\cite{Haxton96}
(illustrated in  their Figure~1) is  grossly
inconsistent with the observed helioseismological measurements.
The vertical scale of Figure~\ref{fig:two} had to be 
expanded by a factor of $2.5$ relative to Figure~\ref{fig:one} in order 
to display the large discrepancies with observations for the mixed model.
The discrepancies for the CH mixed model
(dashed line in Figure~\ref{fig:two})
range from $+8$\% to $-5$\%.
Since $\mu$ in a standard solar model 
decreases monotonically outward from
the solar interior, the mixed model--with a constant value of $\mu$--
predicts too large values for the sound speed in the inner mixed
region and too small values in the outer mixed region.
The asymmetric form of the discrepancies for the CH model is due
to the competition between the assumed constant rescaling of the
temperature in the BP No Diffusion model and the assumed 
 mixing of the
solar core (constant value of $\mu$).
We also show in Figure~\ref{fig:two} the relatively
tiny discrepancies found for the new standard
model, OPAL EOS.   

More generally, 
helioseismology rules out all solar models with large amounts of
interior mixing, unless finely-tuned compensating changes in the
temperature are made.  The mean molecular weight
in the standard solar model with diffusion varies monotonically 
from $0.86$ in the
deep interior to $0.62$ at the outer region of nuclear fusion 
($R = 
0.25 R\odot$) to $0.60$ near the solar surface.
Any mixing
model will cause $\mu$ to be constant and equal to
the average value in the mixed region.
At the very least, 
the region in which nuclear fusion occurs must  be mixed 
in order to 
affect significantly the calculated neutrino 
fluxes\cite{Bahcall89,EC68,BBU68,SS68,Schatzman69}.
Unless almost precisely canceling temperature changes are assumed,
solar models in which the nuclear burning region is mixed ($R \lesssim
0.25 R_{\odot}$)
will give maximum  differences, $\delta c$, between
the mixed and the 
standard model predictions, and hence between the mixed model
predictions and the observations, of order
\begin{equation}
{\delta c \over c} ~=~
{1 \over 2} \left({ {\mu - < \mu >} \over {\mu} }\right) ~\sim~ 7\%
~{\rm to}~ 10\%,
\label{maximum}
\end{equation}
which is inconsistent with Figure~\ref{fig:one}.

Are the helioseismological measurements sensitive to the rates of the
nuclear fusion reactions?  In order to answer this question in its
most extreme form, we have
computed a  model in which the cross section factor, $S_{34}$,
 for the 
$^3{\rm He}(\alpha, \gamma)^7{\rm Be}$ 
reaction is artificially set equal to zero. 
The neutrino fluxes computed from this unrealistic model 
have been  used\cite{Bahcall89}
  to set a lower limit on the allowed 
rate of solar neutrinos in the gallium experiments if
the solar luminosity is currently powered by nuclear fusion reactions. 
Figure~\ref{fig:two} shows that 
although the maximum 
discrepancies ($\sim 1$\%) for  the $S_{34} = 0$ model
are much smaller than for mixed models,
they are still large compared to the differences between  the standard
model and 
helioseismological measurements.
The mean squared discrepancy for the $S_{34} = 0$ model is 19 times
larger than for the standard OPAL EOS model.
We conclude that the $S_{34} = 0$ model is not compatible with
helioseismological observations (see also Ref. \cite{Dziemb94}).

Some  nuclear parameters 
are important for solar neutrino experiments but have  negligible 
effects on  the computed solar model values of the sound
speed. For example, we computed a standard solar model 
in which we artificially decreased by a factor of two 
 the crucial cross
section factor, $S_{17}$,  for the rare 
$^7{\rm Be}(p, \gamma)^8$B reaction.
The sound speeds computed for  this radically different value of
$S_{17}$ differ by less than 1 part in $10^4$ 
from the standard model values.

Finally, we comment on the effects of the recent improvements in
opacity\cite{Opacity} and equation of state\cite{eos} on the predicted
solar neutrino fluxes.  Table~\ref{fluxes} gives the neutrino fluxes
computed for a series
of three different standard solar models, all of which include helium
and heavy element diffusion.  The model labeled BP95 is
from \cite{BP95}; the models labeled New Opac and OPAL
 EOS include,
respectively, the improved opacities discussed in \cite{Opacity} and
the improved opacities plus the new OPAL equation of state discussed in
\cite{eos}.   
\begin{table}[htb]
\centering
\caption[]{Neutrino Fluxes for Solar Models with Diffusion.  All
fluxes, except for $^8$B and $^{17}$F, 
 are given in units of $10^{10}$ per ${\rm cm^{-2}s^{-1}}$ at
the earth's surface. The $^8$B and $^{17}$F fluxes are in units of 
$10^{6}$ per ${\rm cm^{-2}s^{-1}}$.}
\begin{tabular}{lccccccc}
\multicolumn{1}{c}{Model}&$pp$&$pep$&${\rm ^7Be}$&${\rm ^8B}$&${\rm
^{13}N}$&${\rm ^{15}O}$&${\rm ^{17}F}$\\
\noalign{\medskip}
\hline
\noalign{\medskip}
BP95&5.91&0.014&0.515&6.62&0.062&0.055&6.48\\
New Opac&5.91&0.014&0.516&6.62&0.062&0.055&6.48\\
OPAL EOS&5.91&0.014&0.514&6.60&0.062&0.054&6.45\\
\end{tabular}
\label{fluxes}
\end{table}

The neutrino fluxes computed with the improved opacity and equation of
state  differ from the previously published values\cite{BP95} by 
amounts that are negligible in solar neutrino calculations.
The predicted event rate, for all three models, 
is 
\begin{equation}
{\rm Cl~Rate} ~=~9.5^{+1.2}_{-1.4}~~{\rm SNU}
\label{clrate}
\end{equation}
for the chlorine experiment 
and 
\begin{equation}
{\rm Ga~Rate} ~=~137^{+8}_{-7}~~{\rm SNU}
\label{garate}
\end{equation}
for the gallium experiments.
The only noticeable change in the predicted
event rates for the chlorine and the gallium experiment is a 
 2\% larger event rate for chlorine, which is due to a small
improvement\cite{b8cross} 
in the calculation of the neutrino absorption cross
sections for $^8$B.  

We conclude that the recent improvements in opacity and
equation of state do not significantly affect the calculated neutrino
fluxes, although they do result in sound speeds near the solar surface
that are closer to the measured helioseismological values (see
Figure~\ref{fig:one}). 
The calculations of standard solar models lead to predicted
sound speeds that agree  closely
with the measured helioseismological values.
We cannot rule out with mathematical rigor the 
possibility\cite{Antia95} of constructing
nonstandard models, consistent with quantum mechanics and with other
stellar evolution observations, 
that are tuned to give the 
same sound speeds as
the standard solar models.
However, Ockham's razor  suggests a strong preference
for  standard solar models.

We thank P. Demarque for a subroutine that
contains convenient code for  the  OPAL equation of state. 
This work was supported by NSF Grant No. PHY95-13835,
and by the Danish National Research Foundation through the
establishment of the Theoretical Astrophysics Center.

\end{document}